\documentclass[
 reprint,
 superscriptaddress,
 longbibliography,
 amsmath,amssymb,
 prx
]{revtex4-2}
\usepackage{xcolor}

\usepackage[normalem]{ulem}
\usepackage{array}
\usepackage{graphicx}
\usepackage{dcolumn}
\usepackage{bm}

\usepackage{siunitx}

\usepackage{braket}
\usepackage{hyperref}
\usepackage{placeins}

\DeclareSIUnit{\dBm}{dBm}

\begin{document}

\preprint{APS/123-QED}

\title{High-fidelity control of a $^{13}$C nuclear spin coupled to a tin-vacancy center in diamond}

\newcommand{\kitphi}{Physikalisches Institut, Karlsruhe Institute of Technology, Kaiserstraße 12, 76131 Karlsruhe, Germany}

\newcommand{\saarb}{Department of Physics, Saarland University, Campus E2 6, 66123 Saarbrücken, Germany}

\newcommand{\kassel}{Institute of Nanostructure Technologies and Analytics (INA), Center for Interdisciplinary Nanostructure Science and Technology (CINSaT), University of Kassel, Heinrich-Plett-Straße 40, 34132 Kassel, Germany}

\newcommand{\kitiqmt}{Institute for Quantum Materials and Technologies, Karlsruhe Institute of Technology, Kaiserstraße 12, 76131 Karlsruhe, Germany}

\author{Jeremias Resch}
\thanks{These two authors contributed equally}
\affiliation{\kitphi}%

\author{Ioannis Karapatzakis}
\thanks{These two authors contributed equally}
\affiliation{\kitphi}%

\author{Mohamed Elshorbagy}
\affiliation{\kitphi}%

\author{Marcel Schrodin}
\affiliation{\kitphi}%

\author{Philipp Fuchs}
\affiliation{\saarb}%

\author{Philipp Graßhoff}
\affiliation{\kassel}%

\author{Luis Kussi}
\affiliation{\kitphi}%

\author{Christoph Sürgers}
\affiliation{\kitphi}%

\author{Cyril Popov}
\affiliation{\kassel}%

\author{Christoph Becher}
\affiliation{\saarb}%

\author{Wolfgang Wernsdorfer}
\affiliation{\kitphi}
\affiliation{\kitiqmt}%

\author{David Hunger}
\email{david.hunger@kit.edu}
\affiliation{\kitphi}
\affiliation{\kitiqmt}%

\date{\today}

\begin{abstract}
Nuclear spins near group-IV defects in diamond are promising candidates for quantum memories in quantum network applications. Here, we demonstrate high-fidelity control of a single $^{13}$C nuclear spin coupled to a tin-vacancy (SnV) center in diamond. We perform a combination of optical and microwave pumping to achieve initialization into a combined electro-nuclear spin state with a fidelity of $99.74(3)\,\%$. Harnessing a superconducting waveguide for radio-frequency driving, we demonstrate precise nuclear spin control: Ramsey measurements reveal a coherence time of $T_2^* = 1.5(1)\,$milliseconds, and we use dynamical decoupling to extend it to $1.35(3)\,$seconds. We perform randomized benchmarking, yielding a single-qubit gate fidelity of $99.92(1)\,\%$. This demonstrates a coherent spin-photon system with promising properties for quantum network nodes.
\end{abstract}

\maketitle
Nuclear spins in the vicinity of optically addressable spin defects in diamond are promising candidates for spin-photon interfaces, which represent an essential building block for the implementation of quantum networks \cite{Kimble2008}. 
It has been shown that group-IV centers yield coherent optical transitions with minimal spectral diffusion \cite{bradac_quantum_2019,schroder_scalable_2017,trusheim_transform-limited_2020,Görlitz2022,karapatz2024}, offer long-lived electron spins that can be driven coherently \cite{SukachevSiV, Stas2022,Rosenthal, Guo, karapatz2024}, and give access to nuclear spins as quantum memories, harnessing either the defect-centered $^{29}$Si \cite{Stas2022}, $^{73}$Ge \cite{adambukulam_hyperfine_2024}, or $^{117}$Sn \cite{harris_high-fidelity_2025} spins, or neighboring $^{13}$C nuclei \cite{nguyen2019,grimm2025, Beukers2025, klotz2025bipartite}.

The tin-vacancy (SnV) center has recently gained attention due to its bright optical transition \cite{trusheim_transform-limited_2020}, long spin lifetime at elevated temperature, and access to fast spin manipulation in a strained diamond lattice \cite{Rosenthal, Guo, karapatz2024}.
In addition, single-shot readout of the electron spin has been demonstrated \cite{RosenthalSingleShot}. Although the coherence of the electron spin could be extended to $\SI{10}{\milli\second}$ \cite{karapatz2024}, this is still not sufficient for building quantum networks.
Recent work on SiV and GeV centers have shown the possibility to extend the coherence time of a single proximal $^{13}$C nuclear spin up to the seconds range \cite{Stas2022, grimm2025}.
Also, protocols have been demonstrated to control weakly coupled $^{13}$C spins with group-IV defects and their electron spin-1/2 \cite{Beukers2025,klotz2025bipartite}. Still, using such spins as quantum memories requires high-fidelity manipulation, which so far has only been reported for NV centers \cite{bartling_universal_2025}.

In this work, we demonstrate high-fidelity control of a single strongly-coupled $^{13}$C nuclear spin located near a negatively charged SnV center.
We investigate the hyperfine interaction using optically detected magnetic resonance (ODMR), and initialize the nuclear spin into a selectively chosen spin state with a fidelity exceeding $\SI{99}{\percent}$ by employing a combination of optical and microwave (MW) pumping. Using a superconducting coplanar waveguide, we demonstrate direct radio-frequency (RF) control of the nuclear spin with negligible heating, enabling high-fidelity coherent manipulation.
Using Carr–Purcell–Meiboom–Gill (CPMG) decoupling sequences, we extend the nuclear spin coherence time to up to $\SI{1.35\pm0.03}{\second}$, with potential for further extension.
Finally, we determine a single-qubit gate fidelity of $\SI{99.92\pm0.01}{\percent}$ through randomized benchmarking. In addition to the beneficial spin properties we observe excellent optical coherence and spectral stability of the optical transition, which in combination achieves key requirements for a quantum network node.

\begin{figure*}[tb]
\centering
\includegraphics{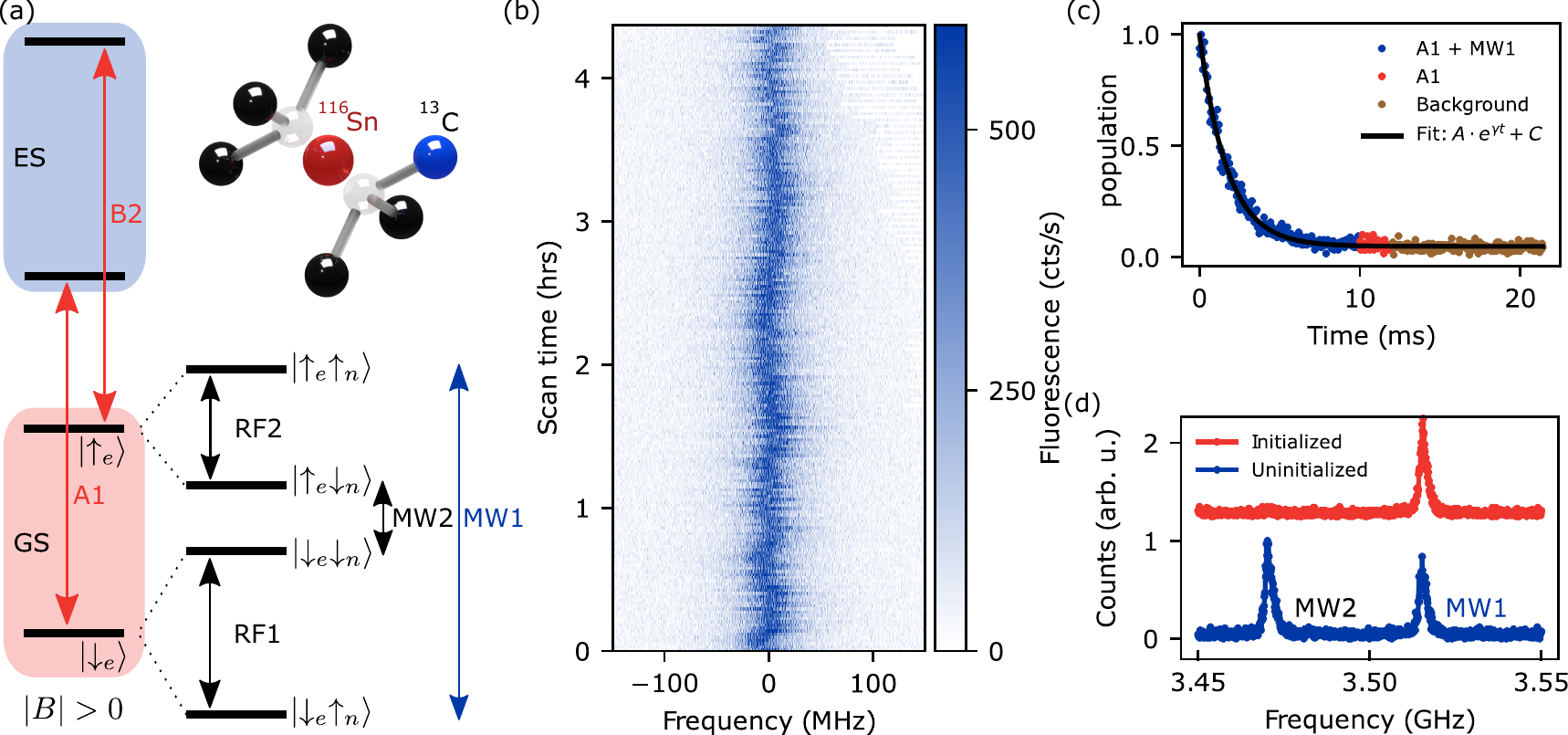}
\caption{\label{fig:Fig1} (a) Energy-level scheme of the SnV center coupled to a proximal $^{13}$C nuclear spin in a non-zero magnetic field. The electron spin levels $\ket{\downarrow_e}, \ket{\uparrow_e}$ are split by the electronic Zeeman effect. Coupling to the $^{13}$C nuclear spin leads to a further splitting into four sublevels. (b) Photoluminescence excitation measurement performed at zero magnetic field. A single sweep linewidth yields a FWHM of $\SI{34\pm4}{\mega\hertz}$. Repeated scans show absence of spectral jumps and charge state switching. (c)~Initialization into the state $\ket{\uparrow_e\downarrow_n}$ by driving the optical transition $\mathrm{A1}$ and the microwave transition $\mathrm{MW1}$ for \SI{10}{\milli\second}. A fit yields an initialization fidelity of $F=99.74(3)\,\%$. (d) ODMR measurement showing the hyperfine splitting of the electron spin transition due to the $^{13}$C nuclear spin, yielding two microwave transitions $\mathrm{MW1}$ and $\mathrm{MW2}$ (blue data), which can be attributed to the two RF transitions shown in (a). After initialization into the state $\ket{\uparrow_e\uparrow_n}$, only one resonance remains (red data).}
\end{figure*}

\section{The setup} This study uses an electronic grade diamond membrane with a natural abundance of \SI{1.1}{\%} $^{13}$C nuclear spins. 
For coherent microwave and radio-frequency spin control, a coplanar waveguide (CPW) made out of niobium with a \SI{5}{\micro\meter} gap at the end of a constriction is fabricated on the diamond by optical lithography and subsequent electron beam evaporation of niobium. For further details on the waveguide fabrication, see \cite{karapatz2024}.
The sample is mounted in a home-built dilution cryostat with a base temperature of \SI{50}{\milli\kelvin} and is surrounded by a 3D vector magnet. Optical access is given through an objective and optical windows towards a confocal microscope setup. 

Fig.~\ref{fig:Fig1}(a) shows a simplified energy-level scheme and the atomic defect structure of the SnV center \cite{Trusheim_2020}. Only the lower orbital branch of the ground (exited) state GS (ES) is shown. The red sphere represents the tin ion on the interstitial site between two vacancies which are depicted as grey spheres. A $^{13}$C nuclear spin is highlighted by a blue sphere in the surrounding carbon lattice.

We probe the system with a narrow-band laser resonantly driving the spin-conserving A1 transition [see Fig.~\ref{fig:Fig1}(a)] and detect the phonon-sideband (PSB) fluorescence. 
Using sufficiently low powers for resonant excitation ($p_\mathrm{PLE}=\SI{0.6}{\nano\watt}\approx p_\mathrm{sat}/50$) enables the optical line to remain stable for hours, maintaining a narrow linewidth and a stable charge state, see Fig.~\ref{fig:Fig1}(b). A Lorentzian fit to a single measurement yields a close to Fourier-limited linewidth of $\SI{34\pm4}{\mega\hertz}$ compared to a linewidth of the averaged spectrum of $\SI{36.6\pm0.1}{\mega\hertz}$ (see Appendix~\ref{sec:app:sample}~Fig.~\ref{fig:App:PLE_scan2}). The small broadening of the hour-long averaged spectrum compared to the single line sweep shows the negligible spectral diffusion present.
In the rare occasion of a charge transition, a weak repump pulse of a few nW power with a  \SI{532}{\nano\meter} laser is used. This allows for a simple measurement sequence without the need for frequent repumping \cite{Rosenthal, Brevoord_2024} or post-selection \cite{RosenthalSingleShot, Herrmann_2024}. Such high stability is a beneficial ingredient for spin-photon entanglement and quantum network applications.

The electronic spin structure of the SnV center in a magnetic field is discussed in detail in the literature \cite{Rosenthal, Guo, Thiering_magneto_optic, karapatz2024}.
By applying an external magnetic field along the quantization axis of the SnV center, the electronic sublevels are split into the two $\{\ket{\downarrow_e}, \ket{\uparrow_e}\}$ spin levels as shown in Fig.~\ref{fig:Fig1}(a).
A strong coupling of the electron spin to a proximal nuclear spin $\ket{\downarrow_n}$ or $\ket{\uparrow_n}$ leads to a further splitting of these levels. This results to a total of four levels in ascending order $\{\ket{\downarrow_e\uparrow_n}, \ket{\downarrow_e\downarrow_n}, \ket{\uparrow_e\downarrow_n}, \ket{\uparrow_e\uparrow_n}\}$ in the electronic ground state.
The depiction of the excited state is simplified to just two levels, as only the spin-conserving optical transitions $\mathrm{A1}$ and $\mathrm{B2}$ are driven by the applied laser pulses.
There are a total of two nuclear-spin-conserving microwave transitions $\mathrm{MW1}$ and $\mathrm{MW2}$, which drive the electron spin, and two electron spin conserving radio-frequency transitions $\mathrm{RF1}$ and $\mathrm{RF2}$ driving the nuclear spin.
Fig.~\ref{fig:Fig1}(d) shows a typical ODMR measurement, where the frequency of the microwave is swept in a chirp over the expected range of the electron spin resonance, while pumping the spin-conserving optical transition $\mathrm{A1}$ and collecting the PSB fluorescence.
The precise electron spin resonance frequencies can be obtained for arbitrary magnetic fields using the Hamiltonian parameters introduced in earlier work \cite{karapatz2024}. Fitting the optical transitions to the Hamiltonian, we obtain a low strain value corresponding to a frequency shift of \SI{41.3\pm 0.8}{\giga\hertz} for the ground state and $\SI{66\pm3}{\giga\hertz}$ for the excited state (see Appendix~\ref{sec:app:ham} for more details.)

\section{Spin initialization}
To use the nuclear spin as a qubit, it must first be initialized into one of its sublevels. This can, for example, be achieved via the implementation of a $\mathrm{SWAP}$ gate 
~\cite{grimm2025}, using resonances in spin echo sequences \cite{nguyen2019}, or using Hartmann-Hahn conditions \cite{Metsch2019}.
However, due to the low electron Rabi frequency of low-strain SnV centers under magnetic fields aligned with the SnV quantization axis~(see Appendix~\ref{sec:app:coh_control}), we opt for an alternative approach based on optical pumping. This technique involves the simultaneous application of one of the microwave transitions and one of the spin-conserving optical transitions, e.g. MW1 in combination with A1 for an initialization into $\ket{\uparrow_e\downarrow_n}$. 
Fig.~\ref{fig:Fig1}(c) illustrates a representative initialization sequence, in which transition A1 is optically pumped while MW1 is applied at a low power of \SI{-16}{dBm}.
The nuclear spin flips randomly, eventually occupying the state $\ket{\uparrow_e\downarrow_n}$, which is off-resonant with both the laser and microwave fields and thus constitutes a dark state. This scheme results in an exponential decay of the PSB fluorescence signal.
To correct for background fluorescence and to initialize any residual electron spin polarization into the correct spin state, a short (\si{\milli\second}) laser-only pulse is typically appended at the end of the sequence.
Fitting the fluorescence decay to an exponential function with amplitude and offset, we extract the initialization fidelity as $F_\mathrm{init} = 99.74(3)\,\%$ by subtracting the offset from the amplitude and normalizing to the total signal~(see Appendix~\ref{sec:App:init})).
Owing to the high cyclicity of this process,
initialization requires about \SIrange{5}{10}{\milli\second}, depending on the applied optical and microwave power. We find a pure electron cyclicity of $\Lambda_e\sim6000$ (see Appendix~\ref{sec:app:cyclicity}).

Using an analogous pumping scheme but applying the transition MW2, one can initialize in the $\ket{\uparrow_e\uparrow_n}$ state, leaving only MW1 as a bright transition in the ODMR measurement as shown in Fig.~\ref{fig:Fig1}(d).

ODMR measurements reveal a splitting of \SI{44.5}{\mega\hertz}, which indicates a proximal nuclear spin \cite{Defo_C13_Coupl_2021}, most probably one of the next neighbors as indicated in Fig.~\ref{fig:Fig1}(a).
The exact positions of the radio-frequency transitions RF1 and RF2 were determined by an optically detected nuclear resonance (ODNR) measurement by continuously pumping the corresponding optical transition B2 or A1 and driving continuously one of the microwave transitions, while chirping the radio-frequency (see Appendix~\ref{sec:app:ham}~Fig.~\ref{fig:App:odmr_comb}).
Due to the close proximity of the nuclear spin, a significant Fermi contact interaction is present, and it is not possible to extract the parallel and perpendicular coupling parameters of the interaction using the secular approximation \cite{grimm2025}.

\begin{figure*}[tb]
\centering
\includegraphics{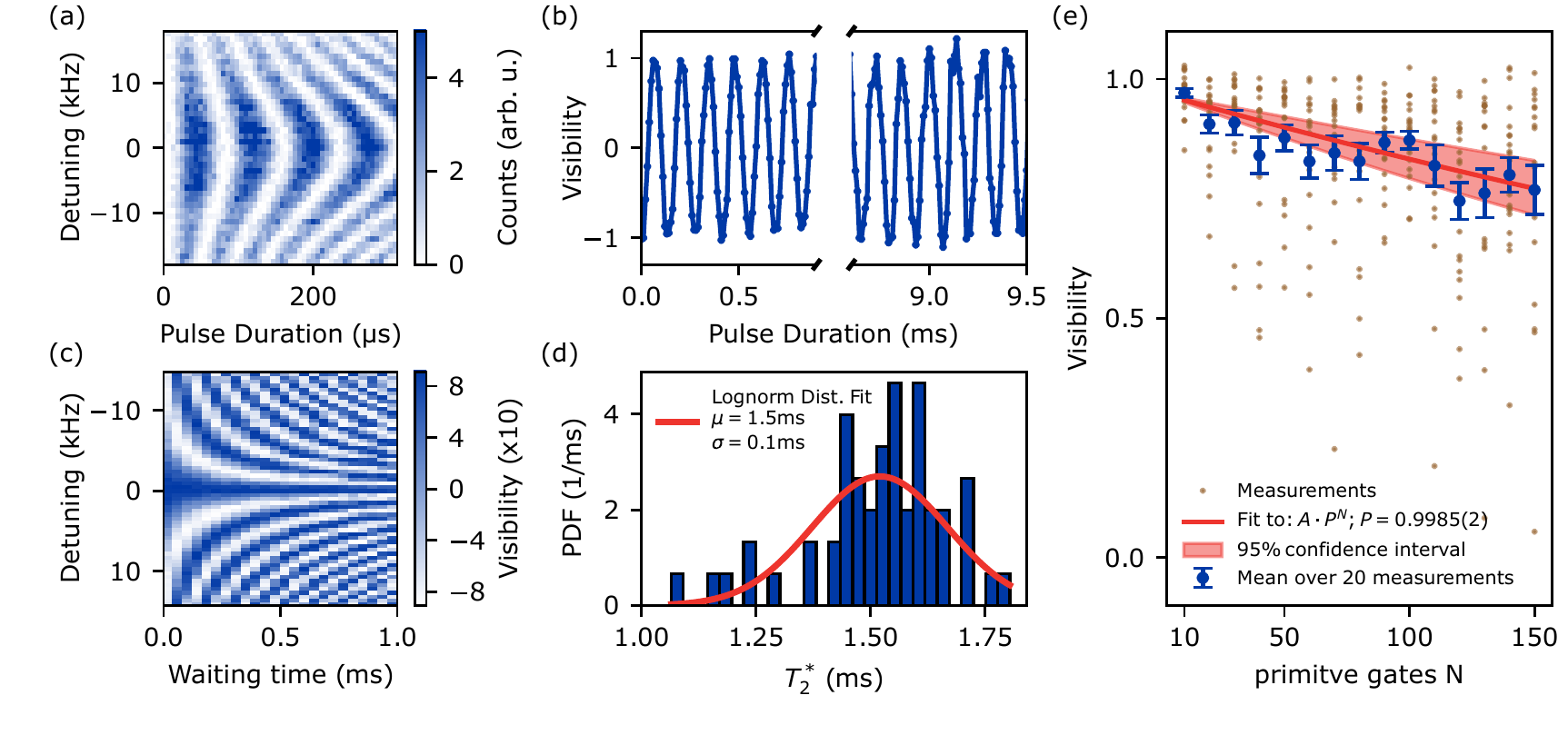}
\caption{\label{fig:Fig2} (a) Chevron pattern of Rabi oscillations for various detunings. (b) Rabi oscillation over a pulse lengths up to \SI{9.5}{\milli\second}. (c) Chevron pattern observed by detuned Ramsey measurements showing the stability of the qubit frequency. (d) Histogram of all coherence times for varying detunings leading to a mean coherence time of $T_2^*=\SI{1.5\pm 0.1}{\milli\second}$. (e) Randomized benchmarking, where $20$ realizations of random gates from the primitive gate set are performed. Each realization is repeated $500$ times, and each blue point on the graph is the mean value of all different realizations. The fidelity per primitive gate is $F=1-(1-P)/2 = 99.92(1)\%$.}
\end{figure*}

\section{Coherent spin control} The ability to initialize the nuclear spin and knowledge of its transition frequency allows for coherent control. We use a superconducting waveguide comparable to the one used for electron spin manipulation in reference \cite{karapatz2024}. Fig.~\ref{fig:Fig2}(a) shows a Rabi chevron pattern around the center frequency \SI{20.998}{\mega\hertz}. This is measured by initializing into the state $\ket{\uparrow_e\uparrow_n}$ via MW2 and coupling to the bright state $\ket{\uparrow_e\downarrow_n}$ by RF pulses at frequency RF2 of varying duration and detuning under a magnetic field of $\SI{60}{\milli\tesla}$ along the SnV axis. With an RF input-power of only \SI{10}{dBm}, a Rabi frequency of \SI{13}{\kilo\hertz} is achieved. The observed nuclear Rabi frequency is slightly enhanced by a factor of $2.07$ compared to a free ${}^{13}$C nuclear spin due to the coupling to the nearby electron spin. For further details on the microwave calibration and estimation of the enhancement, see Appendix~\ref{sec:app:microwave} and \ref{sec:app:nuclearrabi}.
An external magnetic field of \SI{106}{\milli\tesla} was applied to evaluate the quality of the coherent control. The higher field is chosen such that the splitting between the spin-conserving optical lines is larger to avoid cross-pumping of the electron spin.
High-quality resonant Rabi oscillations without reduction of contrast can be observed for $\SI{9.5}{\milli\second}$, see Fig.~\ref{fig:Fig2}(b), which corresponds to an equivalent of $138$ $\pi$-pulses. This demonstrates that the superconducting waveguide can provide extended continuous RF driving without heating, which otherwise represents a challenge for many experiments \cite{Metsch2019,nguyen2019,grimm2025, Stas2022}.
For these long Rabi measurements, we determine the visibility using two Rabi measurements with different pulse lengths, where the second Rabi pulse duration is extended by a single $\pi$-pulse length, resulting in a \SI{180}{\degree} phase shift. The visibility is then calculated as $V=(S_{0}-S_{180})/(S_{0}+S_{180})$, where $S$ indicates the counted fluorescence of the readout pulse on each of the two sequences. %

To measure the coherence and frequency stability of the nuclear spin transition, a Ramsey sequence was performed for various detunings around \SI{20.53}{\mega\hertz}, see Fig.~\ref{fig:Fig2}(c).  A visibility measurement is implemented to account for count rate fluctuations and to give a direct measure of the nuclear spin state population. Therefore, every measurement sequence is repeated with a \SI{180}{\degree} phase shift on the second $\pi/2$-pulse, and the visibility is calculated as above. The full Ramsey chevron measurement is performed over a total measurement time of four days and shows no large frequency jumps of the nuclear transition (see Fig.~\ref{fig:Fig2}(c) and Appendix~\ref{sec:app:coh_control}~Fig.~\ref{fig:app:comparison_fit_data_Ramsey}).
Fitting each oscillation by a sine multiplied with an exponential decay $\exp(-(t/T_2^*)^2)$ gives the decoherence time $T_2^*$.
A histogram of the coherence times for all detunings leads to a mean coherence time of $\bar{T}_2^*=\SI{1.5\pm 0.1}{\milli\second}$, see Fig.~\ref{fig:Fig2}(d).
Note that we can drive Rabi oscillations without any decay even for pulse durations much longer than the $T_2^*$ time [see Fig.~\ref{fig:Fig2}(b)]. This shows that the dephasing originates from slow spectral diffusion.

\section{Randomized Benchmarking}To quantify the achievable fidelity for single-qubit gates, we perform randomized benchmarking (RMBM). We use the set of primitive gates, consisting of the identity operation $I$, \SI{90}{\degree} rotations around the $x$- and $y$-axis of the qubit labeled as $X$ and $Y$ respectively, and \SI{180}{\degree} rotations around the same axis, $X^2$ and $Y^2$. This forms the set $G=\{I, \pm X, \pm Y, \pm X^2, \pm Y^2\}$~\cite{Rosenthal, koch2024}. The identity is defined as a waiting time of the same duration as the $X^2$-pulse to mimic a possible waiting time for a nuclear spin memory readout/transfer.
For each realization of a primitive gate sequence, a random selection out of $G$ is taken. The inverse gate for the whole sequence is found and added to the sequence as a final additional gate \cite{koch2024}.
To obtain a normalized visibility, the sequence is performed twice with two different $\pi$-shifted phases of the final inverse gate. Each visibility measurement uses $20$ random realizations to ensure a statistical distribution of each number of gates used, each repeated $500$ times to collect enough counting statistics.
The mean value of these measurements is marked by blue dots in Fig.~\ref{fig:Fig2}(e).
This mean value can be fitted by the function $A\cdot P^N$, where $A$ is a constant overall fidelity, combining initialization and readout fidelity, $P$ is the fidelity per gate operation, and $N$ is the number of gates \cite{koch2024}. The fidelity per primitive gate is then given by $F_G=1-(1-P)/2=\SI{99.92\pm 0.01}{\%}$. The factor $1/2$ accounts for the gate set being performed twice \cite{koch2024}. This primitive single-qubit gate fidelity can be expressed into the more commonly used single-qubit Clifford gate fidelity \cite{MagesanRMBM_2012, koch2024} of $\SI{99.85\pm0.02}{\%}$.

\section{Dynamical decoupling}
\begin{figure}[tb]
\centering
\includegraphics{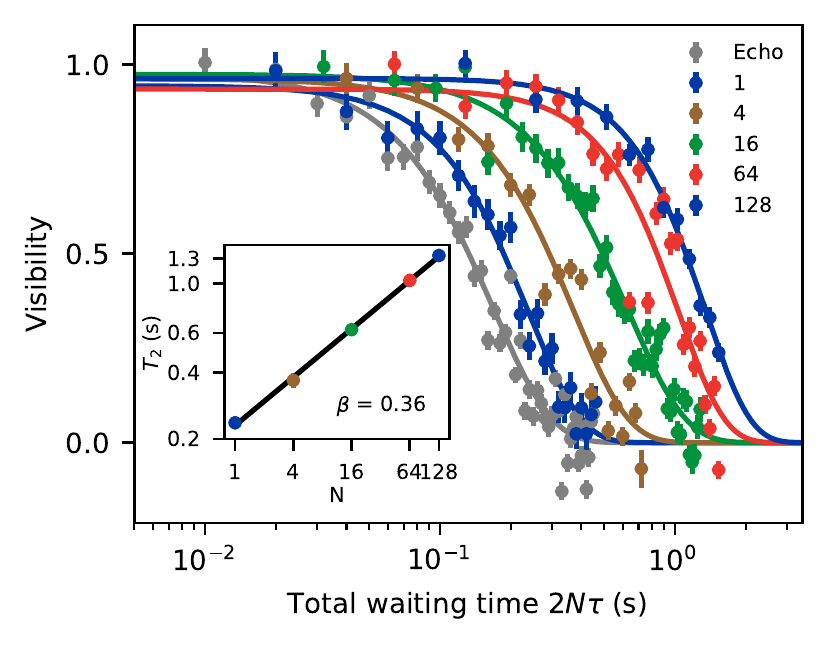}
\caption{\label{fig:Fig3} Extended coherence using dynamical decoupling. A Hahn echo sequence shows a coherence time of $T_2=\SI{167\pm 9}{\milli\second}$ (grey data and fit line). Colored data and fits show dynamical decoupling via a CPMG sequence. The coherence time is increased from initially \SI{233 \pm 8}{\milli\second} to \SI{1.35 \pm 0.03}{\second} using $N=128$ decoupling pulses. The inset shows the scaling of the coherence time $T_2$ with number of pulses $N$.}
\end{figure}

To cancel the influence of slow fluctuations of the surrounding nuclear spin bath, the coherence can be extended with a Hahn-Echo sequence. 
As shown in Fig.~\ref{fig:Fig3} (gray data and fit), the corresponding decay of the visibility follows a stretched exponential decay corresponding to $A \cdot \exp(-(2\tau/T_2)^\xi)$, where $A$ is the initial decoupling fidelity, $T_2$ the coherence time and $\xi$ the stretching factor of the exponential. We find an extended coherence time of $T_2=\SI{167(9)}{\milli\second}$ and $\xi=1.7(2)$.
Assuming an Ornstein-Uhlenbeck model for the bath noise \cite{Wang2012, grimm2025}, we obtain a bath coupling strength of $\SI{930\pm 90}{\hertz}$ from the coherence time $T_2^*$. This coupling strength of the bath is similar to the observed ODNR linewidth of the RF transition, as well as in the same order of magnitude as the detuning distribution in the Ramsey chevron pattern and the observed detuning in the $\SI{9.5}{\milli\second}$ long Rabi measurement (see Appendix \ref{sec:app:nuclearrabi}). In addition, we obtain a correlation time of the bath $\tau_c=\SI{345}{\second}$, which is comparable to recent work on a single $^{13}$C nuclear spin strongly coupled to a GeV center \cite{grimm2025}.

In order to estimate the possible extension of coherence, we perform dynamical decoupling with a Carr-Purcell-Meiboom-Gill (CPMG) sequence. In this sequence, the decoupling $\pi$-pulses are \SI{90}{\degree} phase shifted with respect to the first $\frac{\pi}{2}$-pulse.
For each number of pulses, we vary the delay time $2\tau$ between the pulses and fit the decaying visibility.
For a single decoupling pulse, we obtain a coherence time of $T_2(1)=\SI{233 \pm 8}{\milli\second}$. Increasing the number of decoupling pulses prolongs the coherence time up to $T_2(128)=\SI{1.35 \pm 0.03}{\second}$ using 128 decoupling pulses. This is still not limited by the electron spin lifetime. We find stretching factors ranging from $\xi=1.5$ to $\xi=2.4$ with a mean value $\bar{\xi}=2.0(1)$.
The inset in Fig.~\ref{fig:Fig3}(b) shows the scaling of the coherence time with the number of decoupling pulses, which follows a power law $T_2\propto N^\beta$ with exponent $\beta=0.36$, which is lower than the expected $2/3$ scaling for an Ornstein-Uhlenbeck model of the bath noise. We attribute this to the strong coupling to the electron spin. As we observe a significant improvement between the coherence time $T_2^*$ in the Ramsey decay and the Echo coherence time, we conclude that the estimation of the bath coupling and bath correlation time above are still a good estimation for a slow fluctuating spin bath \cite{senkalla2024, deSousa2009, Wang2012}. We note further that no visibility reduction occurs for increasing pulse number, pointing to the possibility to further extend coherence using more pulses.

\section{Conclusion} We have demonstrated high-fidelity control of a single $^{13}$C nuclear spin utilizing readout via the electron spin of the SnV center. We harness a simple pumping scheme that provides high-fidelity initialization. Employing a superconducting CPW with smaller gap sizes than reported earlier \cite{karapatz2024} enables the demonstration of extended nuclear Rabi oscillations at low driving power with no signs of heating. We use dynamical decoupling to prolong the coherence up to \SI{1.35\pm 0.03}{\second}, reaching a state-of-the-art level as desired for quantum network applications. The control fidelity is quantified by performing randomized benchmarking and achieves values that approach the requirements for quantum error correction \cite{bartling_universal_2025}. In addition, the studied spin is coupled to near-Fourier-limited optical transitions that show only marginal spectral diffusion. Such a combination of high control fidelity and long coherence time of a nuclear spin qubit together with coherent and stable optical transitions fulfills key requirements for spin-photon interfaces and substantiates the potential of SnV centers for quantum network applications.\\

\section*{ACKNOWLEDGEMENTS}
We thank Michael Kieschnick and Jan Meijer for implanting the diamond sample.
We thank Vadim Vorobyov, Katharina Senkalla, Fedor Jelezko, Matthias Müller and Florian Ferlemann for insightful discussions. We thank Thomas Koch and Viktor Adam for helpful discussions and providing scripts for the randomized benchmarking measurements as well as Julia Heupel for technological support.\\

This work was partly supported by the the German Federal Ministry of Research, Technology and Space (Bundesministerium für Forschung, Technologie und Raumfahrt, BMFTR) within the project QR.N (Contracts No. 16KIS2186, No. 16KIS2204, No. 16KIS2180K), QR.X (Contracts No. 16KISQ004, No. 16KISQ001K), SPINNING (Contract No. 13N16211), the Deutsche Forschungsgemeinschaft (DFG) through TRR 306  - 429529648 "QuCoLiMa", the Max Planck School of Photonics (MPSP), and the Karlsruhe School of Optics and Photonics (KSOP).

\appendix 
\section{EXPERIMENTAL DETAILS}
\subsection{Sample fabrication and setup}
\label{sec:app:sample}
The diamond sample used in this paper is cut in the $\langle110\rangle$ directions with a $(100)$ surface with respect to the optical axis. The diamond is etched down along the $(100)$ surface to a total thickness of \SI{26}{\micro\meter} to relieve strain remaining from the manufacturing polishing and reducing the surface roughness \cite{Heupeletch}.
The SnV centers are created by implantation of tin ${}^{116}\mathrm{Sn}$ ions at an energy of \SI{65}{\kilo\electronvolt} into the diamond, and a subsequent annealing of \SI{1200}{\degree C} for \SI{4}{hrs}. For further details on the membrane fabrication, see \cite{karapatz2024, Heupeletch}. Cleaning is performed by boiling with a 1:1:1 mixture of nitric, sulfuric, and perchloric acid (tri-acid cleaning) and subsequent piranha boiling.
The diamond is glued with a thin layer of UV-curing adhesive \textit{(NOA63)} onto an undoped silicon wafer to mitigate background fluorescence. The layer of UV curing glue was chosen to be much thinner compared to previous measurements in reference \cite{karapatz2024} to introduce less strain and achieve higher optical cyclicity \cite{Rosenthal}. The sample is mounted on a copper cold-finger in a home-built dilution cryostat with a base temperature of \SI{50}{\milli\kelvin}, surrounded by a 3D vector magnet. Optical access is given through an objective (\textit{Olympus MPlanN} 100x/0.9NA)
and optical windows towards a confocal microscope setup as described in detail in our previous work \cite{karapatz2024}.

\subsection{Transition frequencies of the electron-nuclear coupled system}
\label{sec:app:ham}
\begin{figure}[tb]
\centering
    \includegraphics[]{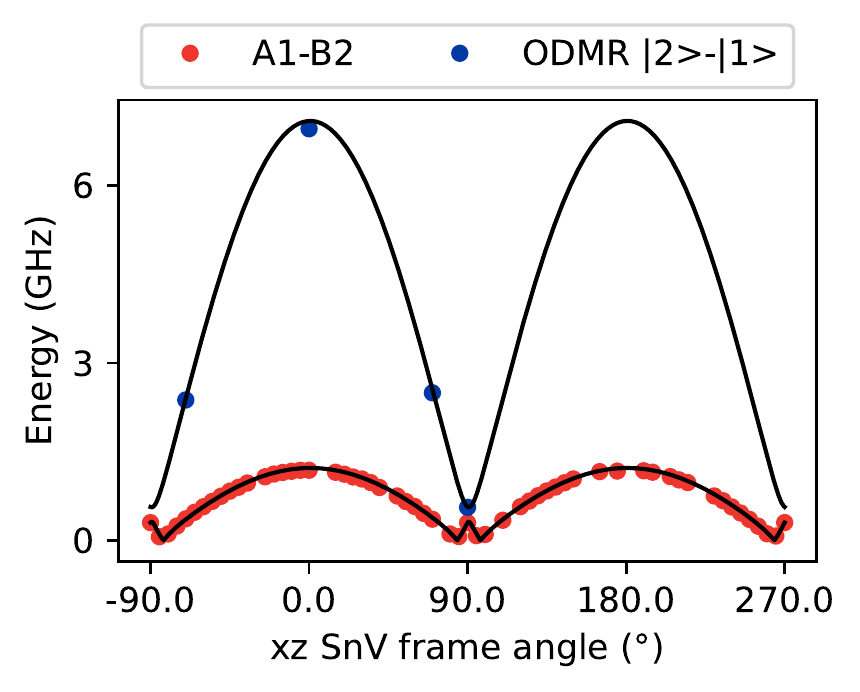}
    \caption{\label{fig:App:Ham_fit} Fit of the electronic energies of the SnV center to the observed optical spin conserving transitions A1 and B2 in the SnV frame (red points). The qubit transition frequencies (blue points) are measured in ODMR scans at 50 mK.}
\end{figure}

The electron spin Hamiltonian can be found by fitting the differences in spin-conserving transitions as well as the ODMR frequencies to the Eigenvalues of the Hamiltonian (see \cite{karapatz2024}).
We find a low strain of $\SI{41.3\pm0.8}{\giga\hertz}$ in the ground state and a strain of $\SI{65.5\pm3.4}{\giga\hertz}$ in the excited state as shown in Fig.~\ref{fig:App:Ham_fit}. Table \ref{tab:App:Hamfit} summarizes the fit parameters.
Using this fit value allows for precise estimation of all electron transition frequencies for arbitrary orientations or magnitudes, reducing measurement overhead significantly.

\begin{figure*}[tb]
\includegraphics[]{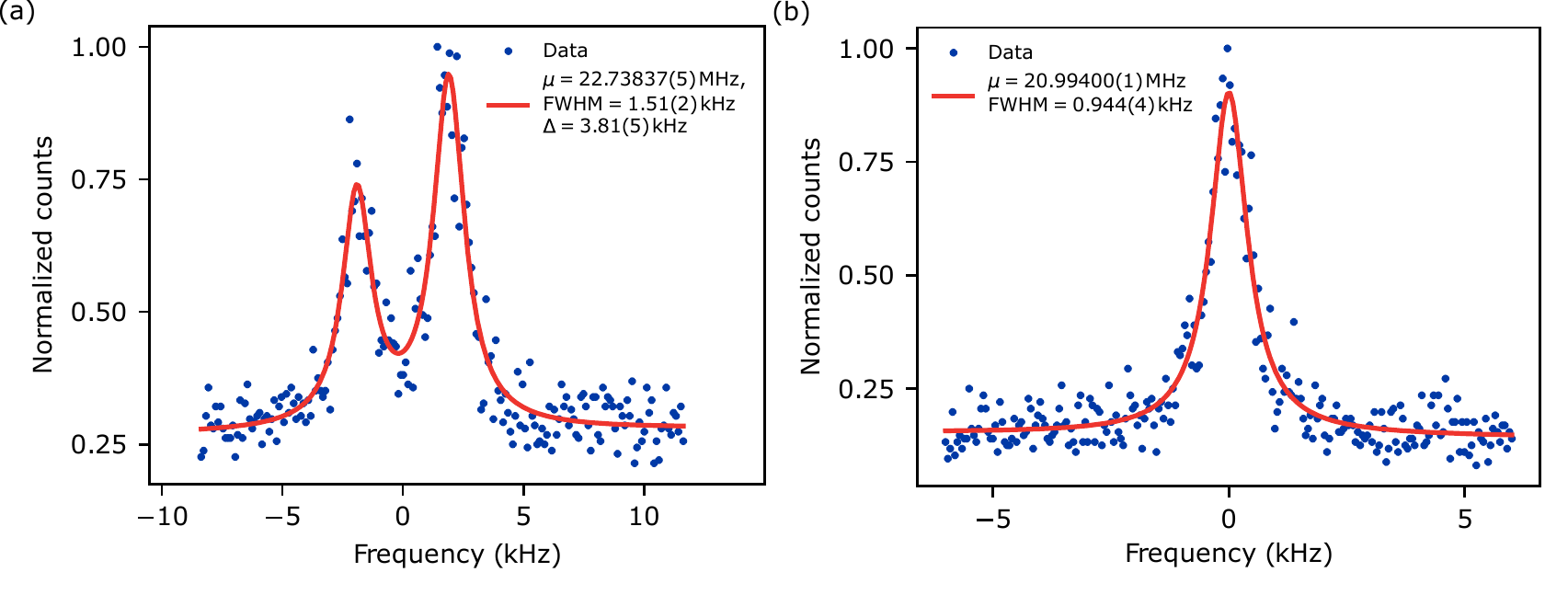}
\caption{\label{fig:App:odmr_comb} (a) Transition frequency of the nuclear hyperfine transition RF1 at a dc magnetic field of $\SI{60}{\milli\tesla}$. Measured via a chirp over the resonance with $\SI{-23}{dBm}$, while continuously pumping the MW1 transition with $\SI{-19}{dBm}$ and the spin-conserving optical transition B2. A fit to two Lorentzian yields a FWHM of $\SI{1.1\pm0.02}{\kilo\hertz}$ and  a splitting $\Delta = \SI{3.81\pm 0.05}{\kilo\hertz}$. (b) Transition frequency RF2 measured in a similar way by pumping the A1 transition. A fit to a single Lorentzian yields a FWHM of $\SI{0.944\pm0.04}{\kilo\hertz}$ in good agreement with the coherence time $T_2^*$ shown in the main text.}
\end{figure*}

The ODNR transitions RF1 and RF2 are shown in Fig.~\ref{fig:App:odmr_comb} for a magnetic field of $\SI{60}{\milli\tesla}$ for qualitative purposes. The ODNR measurement is performed via a chirp over the resonance with $\SI{-23}{dBm}$ RF power, while continuously pumping the MW1 transition with $\SI{-19}{dBm}$ and the spin-conserving optical transition B2/A1 for RF1/RF2, respectively. For RF1, a fit to two Lorentzian yields a FWHM of $\SI{1.1\pm0.02}{\kilo\hertz}$ and  a splitting $\Delta = \SI{3.81\pm 0.05}{\kilo\hertz}$, indicating coupling to a second $^{13}$C nuclear spin. The transition RF2 yields a single Lorentzian. Therefore all experiments are performed on the RF2 transition, where no coupling to a second nuclear spin is observed. A fit yields a FWHM of $\SI{0.944\pm0.04}{\kilo\hertz}$ in good agreement with the coherence time $T_2^*$ shown in the main text.

\begin{table}[tb]
 \caption{\label{tab:App:Hamfit}Fitted strain for the ground and excited state with the relevant Hamiltonian parameters taken from~\cite{karapatz2024}.}
    \centering
    \begin{tabular}{l|c|c}
        \hline
        \hline
        \textbf{Parameter} & \textbf{Value} & \textbf{Source} \\
        \hline
        $\lambda^{\mathrm{g}}$ (GHz) & 822 & \cite{karapatz2024} \\
        $\Upsilon^{\mathrm{g}}$ (GHz) & \SI{41.3\pm 0.8}{} & fitted \\
        $\lambda^{\mathrm{e}}$ (GHz) & 3000 & \cite{Thiering_magneto_optic, Görlitz_2020} \\
        $\Upsilon^{\mathrm{e}}$ (GHz) & \SI{65.5\pm 3.4}{} & fitted \\
        $f_{12}^g$ & 0.251 & \cite{karapatz2024} \\
        $f_{32}^g$ & 0.268 & \cite{karapatz2024} \\
        $f_{12}^e$ & 0.5 & \cite{karapatz2024} \\
        $f_{32}^e$ & 0.486 & \cite{karapatz2024} \\
        $B_\parallel$ (mT) & 200 & fixed \\
        $B_\perp$ (mT) & 200 & fixed \\
        \hline
        \hline
    \end{tabular}
   
\end{table}

\subsection{Microwave power characterization}
\label{sec:app:microwave}

\begin{figure*}[tb]
\includegraphics[]{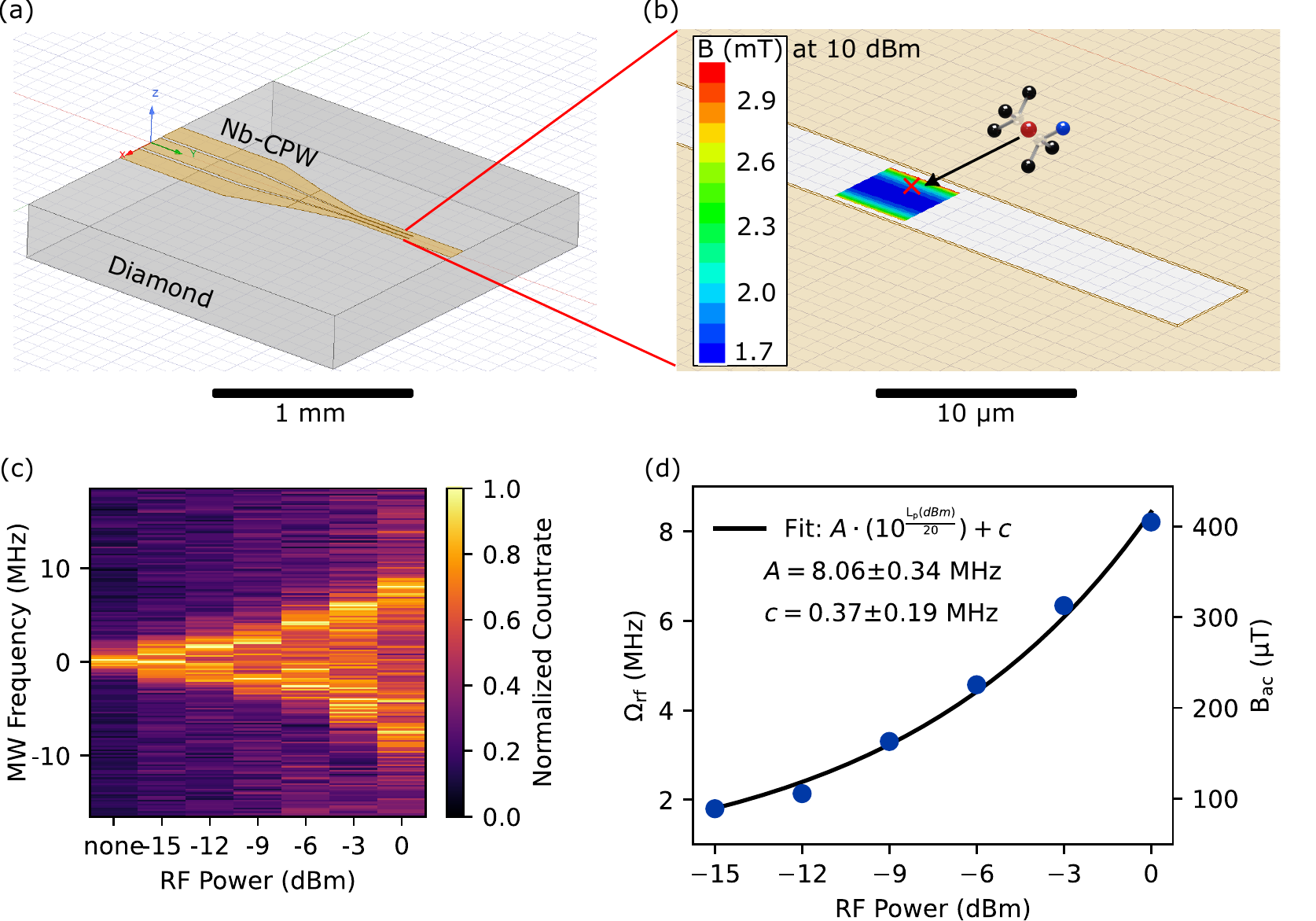}
\caption{\label{fig:App:Sim_HFSS_CPW} (a) Model of the niobium CPW on a diamond substrate created in Ansys HFSS. (b) Simulation of the magnetic field strength of the alternating microwave signal at \SI{10}{\dBm} input power into the structure. The simulation yields a field of $\SI{1.7}{\milli\tesla}$ parallel to the z-lab-axis. (c) Microwave spectrum of the electron transition versus the applied RF-power under two-tone driving. (d) Amplitude of the frequency modulation as a function of the applied RF-power}
\end{figure*}

The microwave transmission setup inside the cryostat is described in~\cite{karapatz2024}.
The only difference in the interior setup is the superconducting coplanar waveguide made of niobium that is used for this work, which is a newly fabricated CPW with a constriction and narrow gaps at the end of the shorted circuit. The CPW has a conductor width of \SI{30}{\micro\meter} at the shorted end and a gap width of \SI{5}{\micro\meter}.
A 3D model of the niobium CPW, created in Ansys HFSS, is depicted in Fig.~\ref{fig:App:Sim_HFSS_CPW}(a).
The narrow gaps confine the alternating magnetic field $B_\text{ac}$ within, allowing fast qubit control at comparable low microwave input powers. In the following, we want to discuss the magnetic field amplitude of the MW/RF fields at the location of the SnV center in detail.

\begin{figure*}[tb]
\includegraphics[]{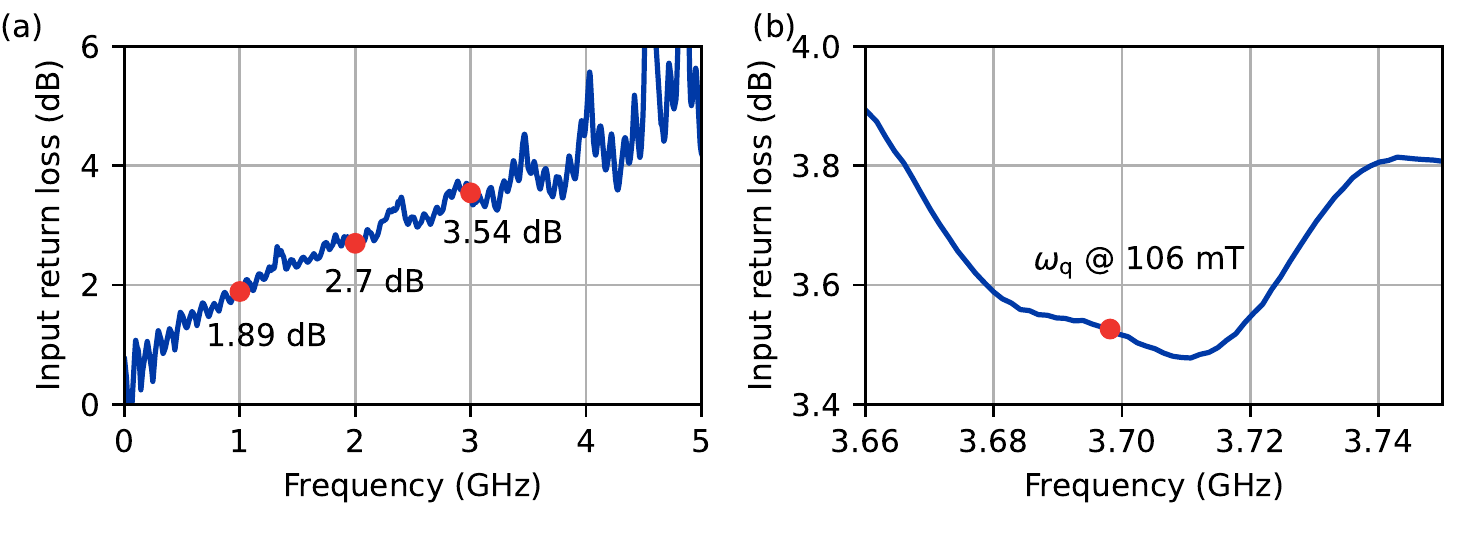}
\caption{\label{fig:App:IL_narrow_CPW} (a) Input return loss of the microwave lines and the sample measured at the input port of the cryostat. (b) The transition of the electron spin resonance is tuned to the local minimum of the input return loss by adjusting the magnetic dc field.}
\end{figure*}

For all measurement configurations, the noted power values correspond to the instantaneous power derived from the MW/RF amplitude and are determined using a spectrum analyzer (Keysight N9323C) in front of the input to the cryostat.
This way, passive elements, like the MW/RF-filters, the attenuators, the combiner and the coaxial cables have no influence.
The input return loss (S11) of the microwave lines and the sample inside the cryostat is depicted in Fig.~\ref{fig:App:IL_narrow_CPW}. A local return loss minimum is located at $\SI{3.7}{\giga\hertz}$, which corresponds to the splitting of the electron spin sublevels for a magnetic field of $\SI{106}{\milli\tesla}$ parallel to the SnV center axis. The loss up to the location of the SnV center is given by roughly half of the corresponding value.

For most of the coherent control measurements, we use a MW/RF input power of about \SI{10}{\dBm}. In the following, we characterize the magnetic field magnitude of the CPW for this value.
In Fig.~\ref{fig:App:Sim_HFSS_CPW}(b) the simulated magnetic field magnitude is depicted within the region of interest for a frequency of \SI{20}{\mega\hertz} and \SI{10}{\dBm} excitation at the input of the structure. We note that the field intensity varies less than one percent at frequencies of \SI{3}{\giga\hertz} according to frequency sweep simulations.
At the center of the CPW gap, the simulation yields an $B_\text{ac}$-field magnitude of about \SI{1.7}{\milli\tesla}. The direction of the $B_\text{ac}$-field at the surface of the gap is oriented perpendicular to the diamond surface and since the diamond is only negligibly misaligned to the \textit{z}-axis of the lab frame ($\sim\SI{0.1}{\degree}$), as spanned by the dc-magnetic field coils, the $B_\text{ac}$-field shows purely in \textit{z}-direction of the lab frame.
For readers interested in the electromagnetic field properties of CPWs, we refer to the work of R. Simons~\cite{CPW_wiley}.
The SnV center is in close proximity to the surface ($\sim\SI{20}{\nano\meter}$), therefore, we assume the same $B_\text{ac}$-field direction at its location.

To get a quantitative analysis of the $B_\text{ac}$-field magnitude, we conduct a two-tone spin resonance excitation of the SnV center. We excite the electronic spin transitions with a weak microwave signal and simultaneously apply a strong RF-signal at a very low frequency of \SI{10}{\kilo\hertz} to modulate the qubit transition.
In the rotating frame of the qubit, the slowly changing $B_\text{RF}$-field can be described as an approximately constant component.
Hence, while probing the spin transition, the energy splitting of the qubit $\omega_\text{q}$ takes a random value between $\omega_\text{q} - \Omega_\text{RF}$ and $\omega_\text{q} + \Omega_\text{RF}$ with an arcsine probability distribution.
Here, $\Omega_\text{RF}$ refers to the maximum magnitude of the sinusoidal qubit transition modulation.

Following the approach in \cite{childress2010, Meinel2021}, adapted for our situation, the Hamiltonian in the rotating wave approximation takes the form:

\begin{equation}
    H_{\mathrm{RWA}}
    = \frac{1}{2}\,\Delta\omega\,\sigma_{z}
    + \frac{1}{2}\,\Omega_\text{MW}\,\sigma_{x}
    + \Omega_\text{RF}\cos\left(\omega_\text{RF}\,t\right)\,\sigma_{z}\, ,
\end{equation}
where $\Omega_\text{MW}$ is the Rabi frequency and $\Delta \omega = \omega_\text{q} - \omega_\text{MW}$ is the detuning from the qubits resonance frequency.
The Pauli matrix term with $\sigma_x$ drives rotations about the $x$-axis, while a $z$-component is added by the magnitude of the detuning.

The third term is due to the $B_\text{RF}$-field that modulates the qubit transition with the applied frequency $\omega_\text{RF}=\SI{10}{\kilo\hertz}$.
Since we know the exact direction of the RF-field and the orientation of the SnV center in the diamond, we can determine the amplitude of the frequency modulation by $\Omega_\text{RF} = \gamma_\text{SnV}(\SI{54.7}{\degree}) B_\text{RF} = \SI{20.27}{\mega\hertz\per\milli\tesla}B_\text{RF}$.
Here, we use the precise knowledge of the electron spin Hamiltonian for this specific SnV center (see Fig.~\ref{fig:App:Ham_fit} for the corresponding fit) and the resulting gyromagnetic ratio for a magnetic field in $z$-direction in the lab frame.
Fig.~\ref{fig:App:Sim_HFSS_CPW}(c) shows the resulting microwave spectrum versus the applied RF-power. We fit each line of the spectrum to an arc-sine probability distribution convolved with the Lorentzian lineshape~\cite{childress2010}. The splitting scales $\propto A\left(10^{\frac{\text{L}_\text{p}(\text{dBm})}{20}}\right)$ as shown in Fig.~\ref{fig:App:Sim_HFSS_CPW}(d).
For an extrapolated RF-power of \SI{10}{\dBm}, we extract a magnetic field amplitude of $B_\text{ac} \approx \SI{1.26}{\milli\tesla}$, roughly $30\,\%$ less compared to the simulation. 
The deviation can be explained relatively easily. The SMA connectors at the milli-Kelvin plate to the flexible waveguide reflect about \SIrange{0.5}{0.7}{\dB}, and an additional return loss of $\sim \SI{0.4}{\dB}$ occurs at the bonding wires to the CPW on the diamond. These losses can not be extracted from simple S11 measurements as depicted for the setup in Fig.~\ref{fig:App:IL_narrow_CPW} due to the low phase-change of the reflected signals at kHz frequency.
Please note, that the magnetic field component perpendicular to the SnV center's quantization axis that coherently drives the electron spin is given by $B_\text{ac}(\SI{35.3}{\degree}) \approx \SI{1.02}{\milli\tesla}$. For a linearly polarized magnetic field, such as in our case, in the RWA only the co-rotating component with amplitude $B_\text{ac}/2$ drives the spin.

\section{NUCLEAR SPIN INITIALIZATION}
\subsection{Initialization Fidelity}
\label{sec:App:init}
The nuclear initialization is fitted to an exponential decaying curve
\begin{equation}
    A\cdot e^{\gamma t} + C\, ,
\end{equation}
using the python package \textit{lmfit} \cite{lmfit}, where each data point is weighted by $1/\sqrt{N+1}$, where $N$ is the amount of counts per data point, to attribute for Poisson noise statistics. The blue data points in Fig.~\ref{fig:Fig1}(c) in the main text correspond to the initialization while driving the spin-conserving transition A1 and the microwave transition MW1 simultaneously. The red data points mark data taken with only the laser to correct for laser induced background counts. To attribute for constant dark counts $B$ the brown data is taken with everything turned off. To calculate a Fidelity one has to subtract these background counts and calculate
\begin{equation}
     F_\mathrm{init} = 1-\frac{C^\prime}{A^\prime}\, ,
\end{equation}
where $C^\prime=C-B$ amd $A^\prime=A-B$
Evaluating the data in Fig.~1(c) in the main text yields the parameters in Table~\ref{tab:nuclear_init}. To estimate the uncertainty of the initialization fidelity we define the functions $g={A^\prime}^2-C^\prime A^\prime$ and $h={A^\prime}^2$ and the fidelity as $F=g/h$. The uncertainty on $g$, $h$ and $F$ is given by considering the partial derivatives, correlations $\rho$ and uncertainties $\sigma$. Using the fact that a higher background leads to higher counts and thus the correlations $\rho_{AB}=1,\ \rho_{CB}=1$ and the correlation coefficient of the fit $\rho_{AC}=0.2235$. Using the correlation of $\rho_{gh}=1$ and the values from fitted and obtained in Table~\ref{tab:nuclear_init}, we find an initialization Fidelity of
\begin{equation}
    F=\SI{99.74\pm 0.03}{\%}\, ,
\end{equation}
for the dark count corrected case and 
\begin{equation}
    F=\SI{95.16\pm 0.04}{\%}\, ,
\end{equation}
without dark count correction.

\begin{table}[tb]
\caption{\label{tab:nuclear_init}
Parameters of the nuclear initialization Fidelity estimation Fit.}
\begin{ruledtabular}
\begin{tabular}{rrr}
Parameter & Value & Standard deviation \\
\hline
Amplitude $A$ & 176 & 3 \\
Decay $\gamma$ & \SI{1.75}{\per\milli\second} & \SI{0.04}{\per\milli\second} \\
Background counts $C$ & 8.40 & 0.4\\
Dark counts $B$ & 8.08 & 0.01\\
\end{tabular}
\end{ruledtabular}
\end{table}

\subsection{Electron and nuclear cyclicity}
\label{sec:app:cyclicity}
\begin{figure*}[tb]
\includegraphics[]{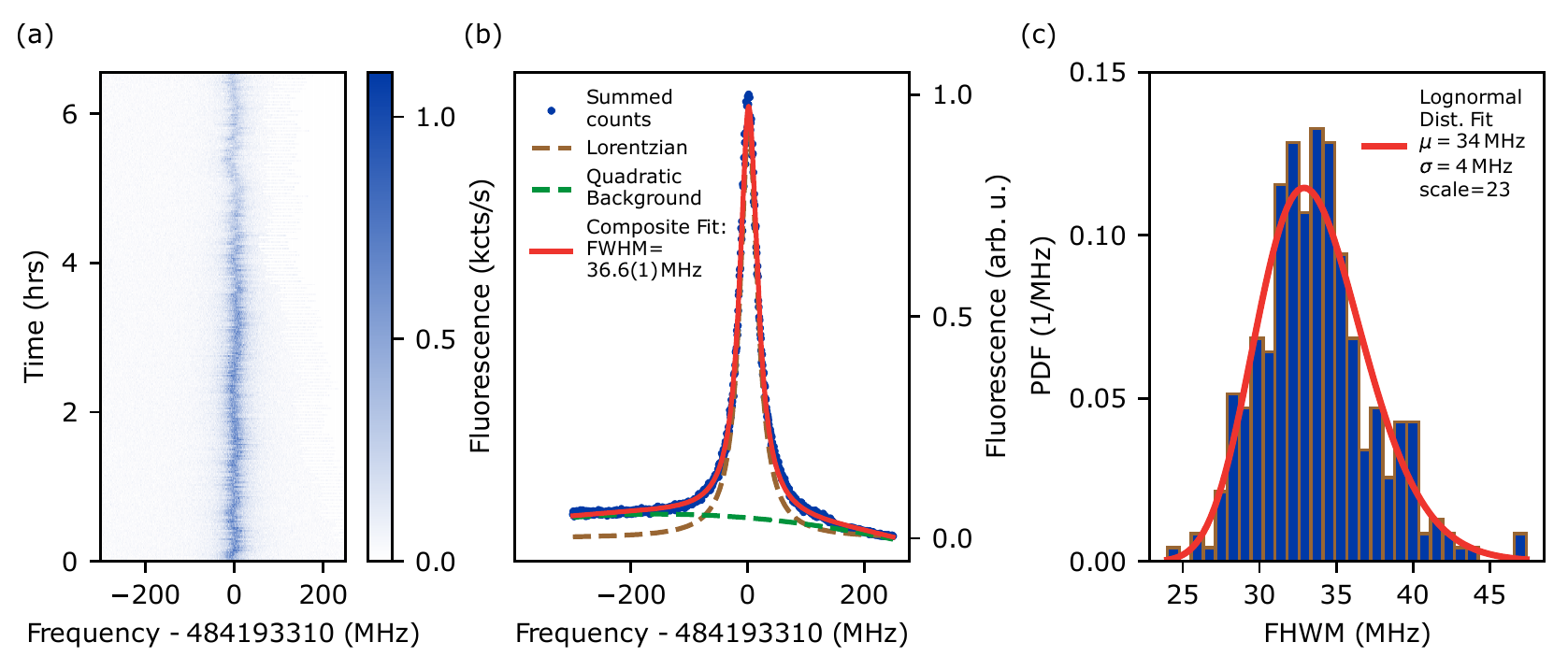}
\caption{\label{fig:App:PLE_scan2} (a) Photoluminescence excitation scan of the optical line A1 with a power of $p_\mathrm{PLE}=\SI{0.6}{\nano\watt}\approx p_\mathrm{sat}/50$. The line remains stable over the course of over six-hour long measurement. (b) Lorentzian Fit to the sum of all individual scans in (a). This leads to a close to Fourier-limited line of $\SI{36.6\pm0.1}{\mega\hertz}$. (c) Individual line fits for each scan result in a linewidth of \SI{34\pm 4}{\mega\hertz} in good agreement with the summed PLE scans showing negligible spectral diffusion.}
\end{figure*}

In order to determine the cyclicity of the nuclear spin initialization all relevant levels have to be considered. We start by a simplified three level system as used in \cite{RosenthalSingleShot, Appel2021} to explain the electron spin initialization dynamics.  Following the derivation in \cite{Appel2021}, we have two spin levels $\ket{\downarrow}$ and $\ket{\uparrow}$ and a excited state $\ket{A}$, which is optically driven by the laser with a optical Rabi frequency $\Omega$. The resulting Hamiltonian is
\begin{equation}
    H=\Delta\ket{A}\bra{A} + \frac{\Omega}{2}\left(\ket{\downarrow}\bra{A}+\ket{A}\bra{\downarrow}\right)\, ,
\end{equation}
where $\Delta$ is the detuning of the laser to the spin conserving transition. Spontaneous emission into the spin ground states is implement by using the rates $\gamma_\mathrm{A\downarrow}$ and $\gamma_\mathrm{A\uparrow}$ respectively. Due to the low temperatures, spin flip rates directly between the two sub levels can be neglected. Following \cite{RosenthalSingleShot}, we introduce a cyclicity $\Lambda_e=P_\mathrm{A\downarrow}/P_\mathrm{A\uparrow}$ given by the ratio of the probability to land in either spin state and the rates can be expressed as $\gamma_\mathrm{A\downarrow}=\gamma\Lambda_e/(1+\Lambda_e)$ and $\gamma_\mathrm{A\uparrow}=\gamma/(1+\Lambda_e)$. Using the assumption that the spin-flipping rate $\gamma_\mathrm{A\uparrow}$ is much smaller than the spin-conserving rate $\gamma_\mathrm{A\downarrow}$, one can set the derivatives of the coherences $\dot{\rho}_\mathrm{A\downarrow}=0,~ \dot{\rho}_\mathrm{\downarrow A}=0$ and end up with the quasi-stationary solutions of the master equation
\begin{align}
    \dot{\rho}_\mathrm{\downarrow\downarrow} &= -W \rho_{\downarrow\downarrow} + \left(W +\gamma_\mathrm{A\downarrow} \right)\rho_\mathrm{AA}\\
    \dot{\rho}_\mathrm{AA} &= W \rho_{\downarrow\downarrow} - \left(W +\gamma \right)\rho_\mathrm{AA}\, , \label{eq:lindblad_electron}
\end{align}
where the rate $W$ is given by $W=\Omega^2\gamma/(4\Delta^2+\gamma^2)$, dependent on the laser detuning $\Delta$ with respect to the spin-conserving transition \cite{Appel2021}. The rate $\gamma$ is given by $2\pi$ times the linewidth of the transition measured via PLE measurements. The resulting linewidth can be fitted by a Lorentzian curve (see Fig.~\ref{fig:App:PLE_scan2}) and results in $\gamma=2\pi\cdot \SI{36.6\pm0.1}{\mega\hertz}=\SI{230.0\pm 0.6}{\mega\hertz}$.

The optical Rabi frequency can be determined from a saturation measurement. For that the resonant power is increased and the count-rate recorded. Fitting to the model
\begin{equation}
    I(p) = I_\mathrm{sat}\cdot \frac{p/p_\mathrm{sat}}{1+p/p_\mathrm{sat}} + n_\mathrm{bgr}\cdot p + C\, ,
\end{equation}
one finds a saturation power of $p_\mathrm{sat}=\SI{29\pm 3}{\nano\watt}$. Using the relation between saturation power and Rabi frequency $p/p_\mathrm{sat} = 2\Omega^2/\gamma^2$ we can convert the used optical power an optical Rabi frequency \cite{foot2005}. 

\begin{figure}[tb]
    \centering
    \includegraphics[]{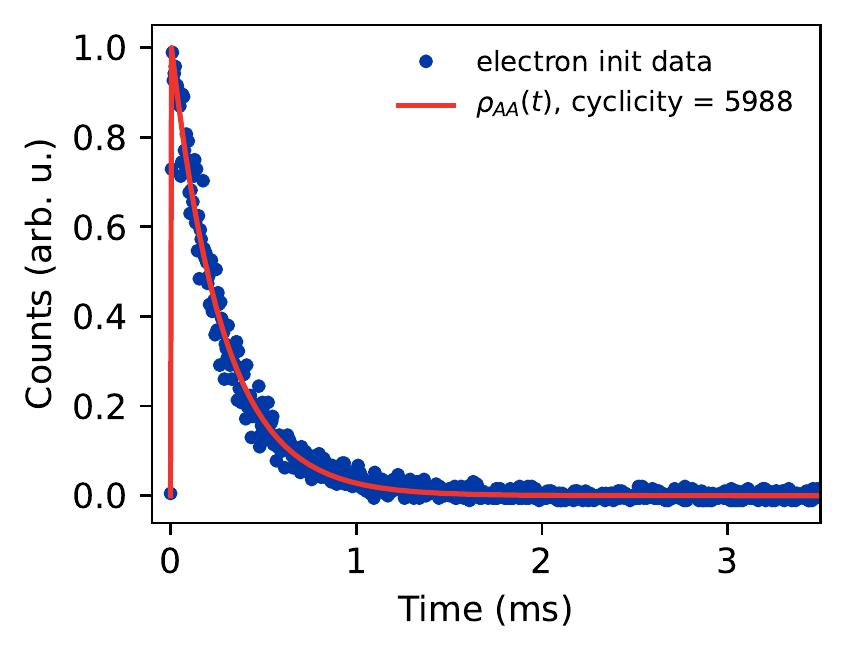}
    \caption{Electron initialization curve measured by pumping the A1 transition. A fit to the rate equation \ref{eq:lindblad_electron} leads to a cyclicity of $\Lambda_e=5988$.}
    \label{fig:App:lindblad_electron_center}
\end{figure}

In order to measure the pure electron initialization we initialize in the state $\ket{\uparrow_e\downarrow_n}$ by pumping for $\SI{10}{\milli\second}$ with microwave MW1 at $\SI{-16}{dBm}$ while continuously pumping the $A1$ transition. Afterwards a $\SI{304}{\nano\second}$ $\pi$-pulse of microwave MW2 at $\SI{17}{dBm}$ flips the electron spin state to $\ket{\downarrow_e\downarrow_n}$ and the initialization is measured by only driving the optical $A1$ transition with $\SI{7}{\nano\watt}$ power. The result is shown in Fig.~\ref{fig:App:lindblad_electron_center}. Assuming a detuning of $\Delta=\mathrm{\omega_{RF1}}=\SI{22.74}{\mega\hertz}$, by setting the $A1$ transition in the center between the two hyperfine transitions, only the cyclicity is left as a free parameter. Fitting the model to the decaying fluorescence in Fig.~\ref{fig:App:lindblad_electron_center} results in a cyclicity $\Lambda_e=5988$. We note that although the exact detuning $\Delta$ is unknown, a change from $\Delta=\SI{0}{\mega\hertz}$ to $\Delta=2\omega_\mathrm{RF1}$ changes the estimated cyclicity only from $6177$ to $5485$, approving the robustness of the electron cyclicity estimation.

In order to estimate the cyclicity of the nuclear-spin-conserving transition, one has to consider all nuclear spin states leading to a total of six relevant levels. By expanding to the nuclear sublevels, we find the populations for ground $g_i$ and excited state $e_i$
\begin{align}
    \dot{g}_i &= \sum_{j=0}^1 (\gamma b_{ji} + R_{ji}) e_j - R_{ji} g_i + W_\mathrm{MW} \left(\xi_{30}\delta_{i0}+ \xi_{03}\delta_{i3}\right)\, ,\\
    \dot{e}_i &= \sum_{i=0}^3 R_{ji} \cdot g_i - (W+\gamma)\cdot e_j\, ,
\end{align}
where the states $g_i$ are $\{\ket{\downarrow_e\uparrow_n}, \ket{\downarrow_e\downarrow_n}, \ket{\uparrow_e\downarrow_n}, \ket{\uparrow_e\uparrow_n}\}$ of the ground state manifold and $e_i=\{\ket{\downarrow_e\uparrow_n}, \ket{\downarrow_e\downarrow_n}\}$ of the excited state in ascending order and $\xi_{ij}=(g_i-g_j)$ the difference between the ground state populations.  The rate $W_\mathrm{MW}$ describes the incoherent microwave pumping with MW1 during the initialization process with $\delta_{ij}$ as the Kronecker-Delta, therefore connecting the ground states $g_0$ and $g_3$.
The matrix $b_{ji}$ gives the probabilities to flip the nuclear and/or electron spin and is defined as 
\begin{equation}
b = \begin{pmatrix}
P_\mathrm{ec}\cdot P_\mathrm{nc} & P_\mathrm{ec}\cdot P_\mathrm{nf} & P_\mathrm{ef}\cdot P_\mathrm{nf} & P_\mathrm{ef}\cdot P_\mathrm{nc}\\
P_\mathrm{ec}\cdot P_\mathrm{nf} & P_\mathrm{ec}\cdot P_\mathrm{nc} & P_\mathrm{ef}\cdot P_\mathrm{nc} & P_\mathrm{ef}\cdot P_\mathrm{nf}
\end{pmatrix}\, ,
\end{equation}
with the probabilities 
\begin{align*}
    P_\mathrm{ec}&=\Lambda_e/(1+\Lambda_e)\, ,\\
    P_\mathrm{ef}&=1/(1+\Lambda_e)\, ,\\ 
    P_\mathrm{nc}&=\Lambda_n/(1+\Lambda_n)\, ,\\
    P_\mathrm{nf}&=1/(1+\Lambda_n)\, .    
\end{align*}
The pump-rates $R$ are given by the matrix
\begin{equation}
   R= \begin{pmatrix}
        W & 0 & 0 & 0\\
        0 & W& 0 & 0
    \end{pmatrix}\,,
\end{equation}
assuming only nuclear and electron spin conserving optical pumping of the transitions.
For these initialization measurements, very low MW powers are used. The resulting Rabi frequency can be calculated by using the electron spin Rabi measurement in Fig.~\ref{fig:App:electron_spin}~(a). For a driving power of $\SI{16}{dBm}$ we observe a Rabi frequency of $\Omega_\mathrm{MW1}=\SI{1.6\pm0.01}{\mega\hertz}$. The resulting attenuated Rabi frequency is given by $\Omega_\mathrm{att} = \Omega_\mathrm{MW1}10^{(\mathrm{att}/20)}$, where a typical attenuation is about $\SI{-35}{dB}$. While the rate model qualitative reproduces the temporal behavior of the initialization for different starting ground states, the incoherent spin pump rate needs to be defined in order to estimate the nuclear cyclicity.
In analogy to the incoherent optical pumping above, we define the incoherent rate as $W_\mathrm{MW}=\Omega_\mathrm{att}^2\gamma_\mathrm{e, sp}/(4\Delta_\mathrm{MW}^2+\gamma_\mathrm{e, sp}^2)$ with the detuning of the microwave $\Delta_\mathrm{MW}$ and the linewidth of the spin transition $\gamma_\mathrm{e, sp}$. The linewidth of the spin transition without any applied optical fields can be approximated by the coherence time $1/T_2^*$, which yields a linewidth of $\gamma_\mathrm{e, sp}=\SI{1.4}{\mega\hertz}$ [see Fig.~\ref{fig:App:electron_spin} (c)].
However measuring the resulting dark spin pump rate is not possible as $\Omega_\mathrm{att}\ll1/T_2^*$, yielding only a contribution to the signal when the electron spin is resonant to the microwave tone, resulting in too small observed rates in the few $\SIrange{10}{100}{\hertz}$ ranges.
Due to the optical pumping during the initialization process, the spin resonance is broadened to the optical Rabi frequency. However, measuring the resulting incoherent spin pump rate under this conditions is not directly accessible. If we use the naive estimation by comparing the $1/e$ decay of electron initialization and nuclear initialization and thus assuming a nuclear cyclicity of $\Lambda_n=10$, we find an effective linewidth $\gamma_\mathrm{e, sp}=\SI{0.49}{\mega\hertz}$, in reasonable agreement with $1/T_2^*$. But due to the strong correlation between nuclear cyclicity and spin pump rate, this does not allow for simultaneous fitting of both to the rate model at the same time. One possibility to estimate the cyclicity are dc magnetic field angle dependent measurements to map out the full hyperfine tensor and estimating the cyclicity via the change of the quantization axis of the nuclear spin between ground and excited state due to the different dipole-dipole coupling~\cite{ulanowski2025cavity}.

\section{COHERENT CONTROL}
\subsection{Nuclear Rabi frequency enhancement}
\label{sec:app:nuclearrabi}

\begin{figure}[tb]
\includegraphics[]{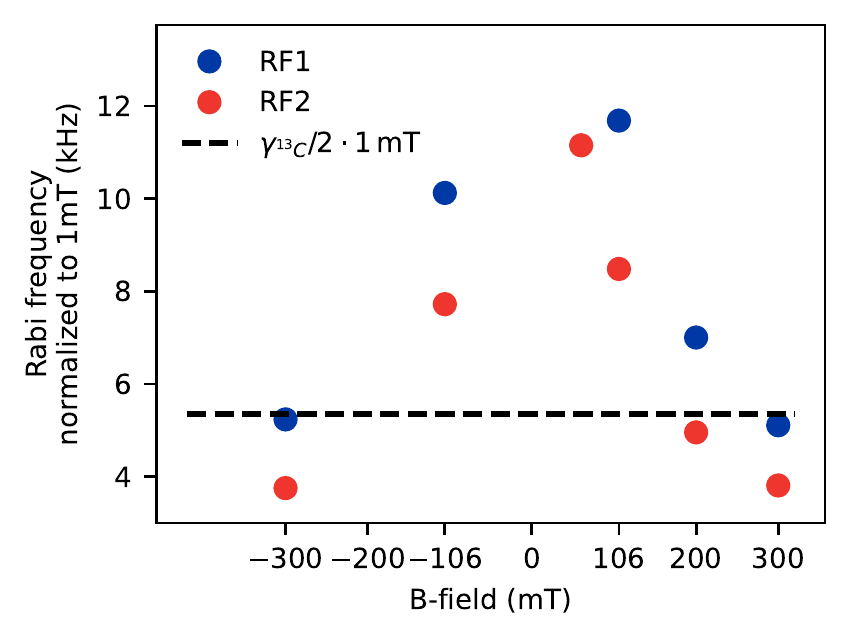}
\caption{\label{fig:App:rabi_vs_bfeld} Nuclear Rabi frequency as a function of the dc magnetic field aligned parallel to the SnV center axis. With increasing amplitude of the dc magnetic field the Rabi frequency decreases. The free Rabi frequency for a linear polarized microwave field perpendicular to ${}^{13}\text{C}$ nuclear spin is denoted by a the black dashed line.}
\end{figure}

Placing a nuclear spin in close proximity to an electron spin enhances the nuclear Rabi frequency by mixing of electronic character into the nuclear spin levels \cite{Smeltzer2009, childress2006}.
To estimate this enhancement of the nuclear Rabi frequency, we measure it as function of the external applied dc magnetic field from $\SI{-300}{\milli\tesla}$ to $\SI{300}{\milli\tesla}$ parallel to the SnV axis. For each dc magnetic field, due to the frequency dependent losses in the microwave setup (e.g. the narrow-band RF-filter), the RF-field $B_\mathrm{ac}$ is calculated using the calibration from the previous section.
A comparison to the gyromagnetic ratio of the ${}^{13}\text{C}$ nuclear spin gives an estimation of the enhancement due to the nearby electron spin \cite{childress2006, Smeltzer2009}. The resulting Rabi frequency is given by 
\begin{equation}
    \Omega_\mathrm{nuc} = \gamma_{{}^{13}\text{C}}/2 \cdot \xi\cdot B_\mathrm{ac}\, ,
\end{equation}
with the gyromagnetic ratio $\gamma_{{}^{13}\text{C}}=\SI{10.7}{\kilo\hertz\per\milli\tesla}$. The factor $1/2$ attributes for the linear polarized microwave field, where only the co-rotating circular polarization is driving the nuclear spin.

The nuclear Rabi frequency depends on the relative angles of the magnetic moment of the ${}^{13}\text{C}$ nuclear spin and the $B_\mathrm{ac}$ field orientation. For small dc fields, the dipole-dipole coupling term dominates and sets the precession axis of the ${}^{13}\text{C}$ nuclear spin. As the dc field increases, the ${}^{13}\text{C}$ precession axis tilts towards the direction of the external dc field, which coincides in this experiment with the quantization axis of the SnV center. Depending on the position of the ${}^{13}\text{C}$ nuclear spin relative to the SnV center, the tilt can coincide with the orientation of the $B_\mathrm{ac}$ field, thus creating a local minimum of the Rabi frequency as the field increases. Only at very strong dc fields the Rabi frequency should approach that of the free ${}^{13}\text{C}$ nuclear spin. 
Thus the measurements of the Rabi frequency in Fig.~\ref{fig:App:rabi_vs_bfeld} are normalized to a field $B_\mathrm{ac}^{\text{SnV},\perp}=\SI{1}{\milli\tesla}$ perpendicular to the SnV axis using the calibration above in \ref{sec:app:microwave} as this is the expected setting giving the driving strength for a free ${}^{13}\text{C}$ nuclear spin under optimal conditions.
We observe a maximum enhancement of $\xi(\SI{60}{\milli\tesla})=2.07$, which yields a nuclear Rabi frequency roughly twofold faster than the expected free Rabi frequency.

\begin{figure}[tb]
\includegraphics[]{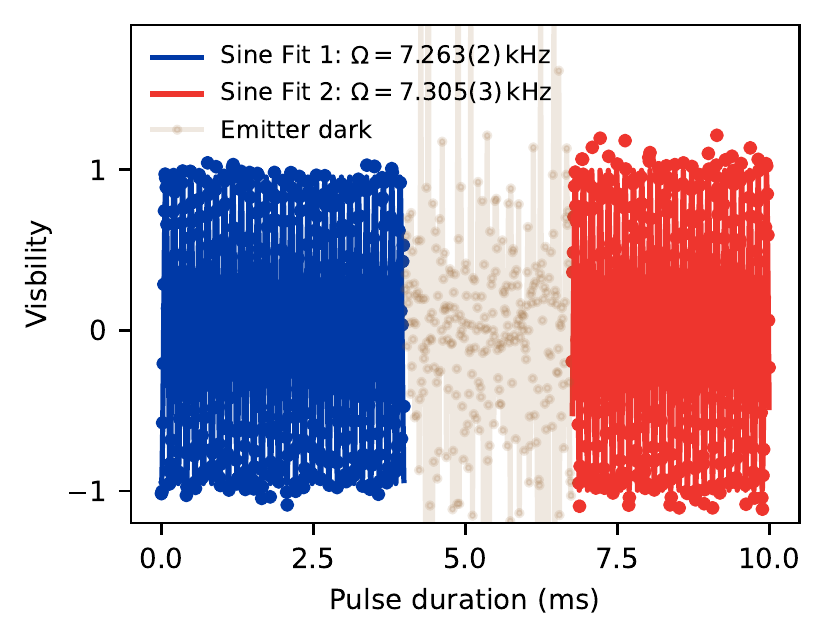}
\caption{\label{fig:App:long_rabi_all} Long Rabi oscillation with intermediate repump. The Rabi oscillation before the repump of $\SI{7.263\pm0.001}{\kilo\hertz}$ only differs slightly from the oscillation after of $\SI{7.305\pm0.003}{\kilo\hertz}$. This would correspond to a detuning of the line of $\SI{826\pm30}{\hertz}$.}
\end{figure}

To quantify the heating effect and loss of coherence due to steady driving, we measure Rabi oscillations up to $\SI{10}{\milli\second}$ pulse length with a power of $\SI{10}{dBm}$ at a field of $\SI{106}{\milli\tesla}$, where the enhancement is $\xi(\SI{106}{\milli\tesla})=1.57$. To compensate for emitter drifts or laser fluctuations during this long measurement, we determine a visibility using two subsequent Rabi measurements, where the second Rabi pulse duration is extended by a single $\pi$-pulse length of $\SI{69.25}{\micro\second}$, resulting in a \SI{180}{\degree} phase shift. The visibility is then calculated as $V=(S_{0}-S_{180})/(S_{0}+S_{180})$, where $S$ indicates the counted fluorescence. The result is shown in Fig.~\ref{fig:App:long_rabi_all}. The emitter lost its charge state during the measurement sequence, but was recovered by a $\SI{532}{\nano\meter}$ laser pulse and showed no loss of coherence afterwards. The change in the fitted Rabi frequency from $\SI{7.263\pm0.002}{\kilo\hertz}$ to $\SI{7.305\pm0.003}{\kilo\hertz}$ can be attributed to detuning of $\Delta=\SI{826\pm30}{\hertz}$. This shows that we can drive the nuclear spin for extended times without any signs of heating. The increase of noise in the visibility can be attributed to a slow spatial drift of emitter and corresponding reduction of fluorescence during the during of the experiment of about $\SI{5}{hrs}$. 

\begin{figure}[tb]
    \centering
    \includegraphics[]{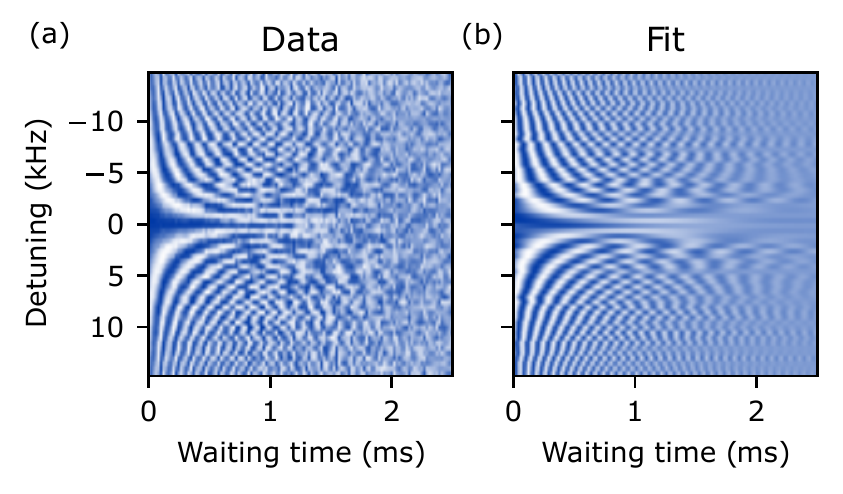}
    \caption{(a) Ramsey measurements of RF2 for varying detunings up to a waiting time of $\SI{2.5}{\milli\second}$. (b) Fitting each row with $\exp(-(t/T_2^*)^2)$ gives the coherence time given in the main text.}
    \label{fig:app:comparison_fit_data_Ramsey}
\end{figure}

\begin{figure}[tb]
\includegraphics[]{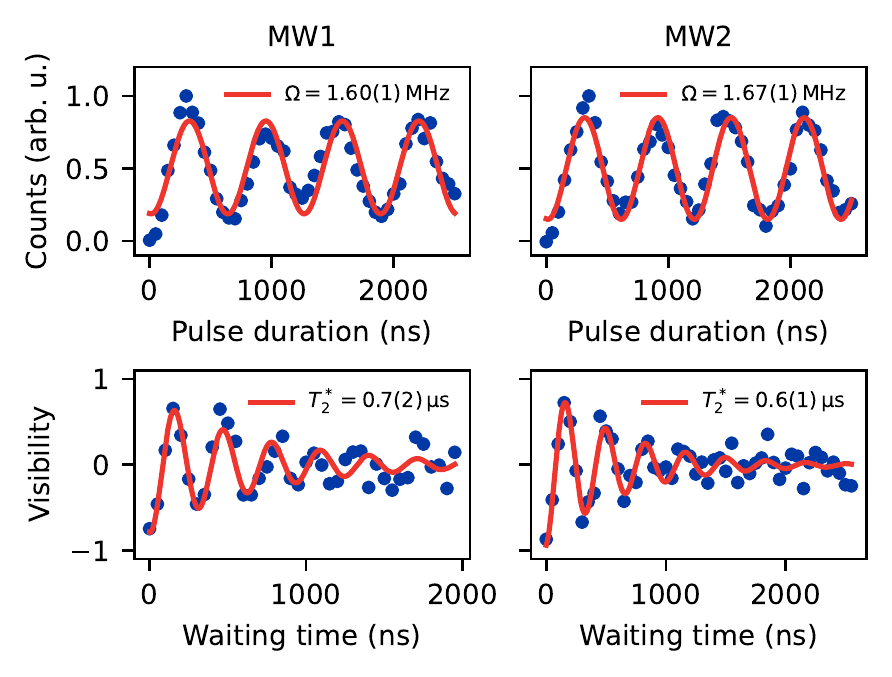}
\caption{\label{fig:App:electron_spin} Electron spin control. (a) Single oscillation with a Rabi frequency of $\Omega=\SI{1.6\pm0.01}{\mega\hertz}$ at $\SI{16}{dBm}$ power for transition frequency MW1. (b) Single oscillation with a Rabi frequency of $\Omega=\SI{1.67\pm0.01}{\mega\hertz}$ at $\SI{14}{dBm}$ MW power for transition frequency MW2. (c) Ramsey decay of MW1 for a detuning of $\SI{3.22\pm0.05}{\mega\hertz}$ yielding a coherence time $T_{2(e)}^*=\SI{0.7\pm0.2}{\micro\second}$. (d) Ramsey decay of MW2 for a detuning of $\SI{2.99\pm0.04}{\mega\hertz}$ yielding a coherence time $T_{2(e)}^*=\SI{0.6\pm0.1}{\micro\second}$ in good agreement with the coherence time of MW1 as expected.}
\end{figure}

In extension to the Ramsey chevron pattern depicted in Fig.~\ref{fig:Fig2}(c), we show here the full Ramsey measurement used for calculating the coherence times given in Fig.~\ref{fig:Fig2}(d). Fitting each row to the expected Gaussian decay $\exp(-(t/T_2^*)^2)$ generates the fits in Fig.~\ref{fig:app:comparison_fit_data_Ramsey}(b), where we can see small deviations of the detuning over the course of the measurement. We note that the total measurement duration was $\sim\SI{100}{\hour}$.

\subsection{Coherent control of the electron spin}
\label{sec:app:coh_control}

For completeness, we show coherent control of the electron spin. 
In order to drive an electronic transition MW2/MW1 we initialize into one of the four sub-states via pumping A1 and MW1/MW2 followed by the application of the wanted measurement sequence of MW2/MW1. Readout is performed again by using the initialization sequence.
For such low-strain emitters the electron Rabi frequency is heavily quenched for a field aligned parallel to the SnV center axis \cite{Pieplow_strain}. Nevertheless, we show coherent Rabi oscillations using the same superconducting waveguide and achieve Rabi frequencies up to $\SI{1.67\pm0.01}{\mega\hertz}$ at input powers of $\SI{14}{dBm}$. To improve Rabi frequencies a finite angle can be chosen \cite{Pieplow_strain, karapatz2024} or slightly more strain induced \cite{Rosenthal, Guo, karapatz2024}. The coherence time measured via Ramsey decay shows slightly worse values of $\SI{0.7\pm0.1}{\micro\second}$, compared to earlier work \cite{karapatz2024, Guo, Beukers2025} but has also been observed before \cite{Rosenthal}.

\FloatBarrier
\bibliography{mybib}
\end{document}